\documentclass[aip,jap,reprint]{revtex4-1}

\usepackage{graphicx}
\usepackage[usenames,dvipsnames]{color}
\usepackage[dvipdfm]{hyperref}
\usepackage[all]{hypcap}
\hypersetup{colorlinks=true,linkcolor=blue,citecolor=blue,urlcolor=blue} 

\newcommand{\mytilde}{\raise.17ex\hbox{$\scriptstyle\mathtt{\sim}$}}

\begin{document}

\title{Electronic band gap reduction and intense luminescence in Co and Mn ion-implanted \texorpdfstring{SiO$_2$}{SiO2}}

\author{R. J. Green}
\email{robert.green@usask.ca}
\affiliation{Department of Physics and Engineering Physics, University of Saskatchewan, 116 Science Place, Saskatoon, Saskatchewan, Canada, S7N 5E2}

\author{D. A. Zatsepin}
\affiliation{Ural Federal University, 19 Mira Str., 620002 Yekaterinburg, Russia}
\affiliation{Institute of Metal Physics, Russian Academy of Sciences-Ural Division, 18 Kovalevskoi Str., 620990 Yekaterinburg, Russia}

\author{D. J. St. Onge}
\affiliation{Department of Physics and Engineering Physics, University of Saskatchewan, 116 Science Place, Saskatoon, Saskatchewan, Canada, S7N 5E2}

\author{E. Z. Kurmaev}
\affiliation{Institute of Metal Physics, Russian Academy of Sciences-Ural Division, 18 Kovalevskoi Str., 620990 Yekaterinburg, Russia}

\author{N. V. Gavrilov}
\affiliation{Institute of Electrophysics, Russian Academy of Sciences-Ural Division, 620016 Yekaterinburg, Russia.}

\author{A. F. Zatsepin}
\affiliation{Ural Federal University, 19 Mira Str., 620002 Yekaterinburg, Russia}

\author{A. Moewes}
\affiliation{Department of Physics and Engineering Physics, University of Saskatchewan, 116 Science Place, Saskatoon, Saskatchewan, Canada, S7N 5E2}

\date{\today}

\begin{abstract}
Cobalt and manganese ions are implanted into SiO$_2$ over a wide range of concentrations. For low concentrations, the Co atoms occupy interstitial locations, coordinated with oxygen, while metallic Co clusters form at higher implantation concentrations. For all concentrations studied here, Mn ions remain in interstitial locations and do not cluster. Using resonant x-ray emission spectroscopy and Anderson impurity model calculations, we determine the strength of the covalent interaction between the interstitial ions and the SiO$_2$ valence band, finding it comparable to Mn and Co monoxides. Further, we find an increasing reduction in the SiO$_2$ electronic band gap for increasing implantation concentration, due primarily to the introduction of Mn- and Co-derived conduction band states. We also observe a strong increase in a band of x-ray stimulated luminescence at 2.75 eV after implantation, attributed to oxygen deficient centers formed during implantation.
\end{abstract}


\maketitle

\section{\label{SEC:Intro}Introduction}

The incorporation of $3d$ transition metal impurities into semiconductors and insulators like ZnO and SiO$_2$ continues to be a highly active research field with many practical applications. Recently, dilute magnetic semiconductors\textemdash materials where transition metal impurities are introduced into semiconductors to induce magnetic properties useful for spintronic \cite{Wolf_Sci_2001} computing\textemdash have garnered copious amounts of interest.\cite{Ogale_AM_2010} Reports of useful ferromagnetic properties continue to appear,\cite{Ogale_AM_2010} though there are still some outstanding questions regarding the origin of magnetism in these materials.\cite{Coey_CTFM_NJP_2010, Coey_NatMat_2005, Ogale_AM_2010} A related field of study involves the incorporation of embedded transition metal nanoclusters into dielectrics. These materials often exhibit interesting and useful properties like nonlinear optical susceptibilities, intense photoluminescence, altered band structures, and superparamagnetism.\cite{Meldrum_AM_2001}  Still another related application of transition metal impurities in semiconductors and insulators is the field of band gap engineering, where electronic band gaps are modified through the introduction of transition metal valence or conduction states upon doping. Such materials are useful for electronics, optoelectronics, and photocatalyst applications, for example.

Many different synthesis approaches can be implemented to obtain the materials described above. Techniques such as pulsed laser deposition, molecular beam epitaxy, sol-gel synthesis, and solid state mixing are often used. Ion implantation is an alternative, very versatile technique. The strengths of ion implantation relate to its high reproducibility, precise locality, and the ability to inject a small, controlled quantity of almost any type of impurity ion into nearly any kind of host material.  Variations of the fluence, or quantity, of certain implanted atoms are known to yield nanoparticles of varying size distributions.\cite{Meldrum_AM_2001}

In this work we study the co-implantation of cobalt and manganese ions into an amorphous silica host matrix. Cobalt is known to form nanoparticles or buried metallic layers in certain conditions,\cite{Cattaruzza_APL_1998} while Mn is more chemically active and consequently interacts more strongly with the host atoms. Thus, dual implantation is expected to produce metallic Co as well as modify the electronic structure of the host SiO$_2$, offering both active nanoparticles and electronic band gap variations which would be useful for applications. Analyzing the implanted materials using x-ray absorption spectroscopy (XAS) and (resonant) x-ray emission spectroscopy [(R)XES], here we develop a detailed description of the crystal and electronic structures for various fluences, as well as study the luminescence properties specific to the implantation region. Our results show varying propensities for metallic aggregation for Mn and Co ions, and show that when in interstitial locations, the ions hybridize with the host valence band to a similar degree as in monoxides. Further, we find a reduction of the SiO$_2$ band gap upon implantation and a strong increase in the x-ray stimulated optical luminescence of the implanted materials, both attributes which are often sought after for applications.

\section{\label{SEC:Exp}Materials and Methods}

The samples studied in this work were prepared from a single block of 99.9\% purity fused quartz glass and consist of square pieces \mbox{$10\times 10\times 1$ mm} in size, with surfaces of optical quality. Before irradiation the samples were washed in an ultrasonic bath of alcohol. Implantation of the Co$^{+}$ and Mn$^{+}$ ions was carried out via the periodic-pulse mode using the metal vapor vacuum arc ion source developed at the Institute of Electrophysics (Yekaterinburg). A powdered arc cathode was made from the sintered mixture of 50\% cobalt and 50\% manganese powders by weight. The operating pressure in the implantation chamber was 1.8-2.0 x 10$^{-4}$ Torr. The ion energy was set to 30 keV, the pulse duration was 0.4 ms, and pulse current density was 0.6 mA/cm$^{2}$. The ion fluence (implantation dose) was \mbox{7 x 10$^{15}$,} \mbox{3 x 10$^{16}$,} and \mbox{2 x 10$^{17}$ cm$^{-2}$} for three different samples prepared (hereafter referred to as samples L, M, and H, respectively for low, medium, and high fluence). During implantation the sample temperatures did not exceed \mbox{350 $^{\circ}$C.}

The XAS experiments of the O $K$ edge and the Mn and Co $L_{2,3}$ edges in this work, as well as luminescence measurements, were performed using the Spherical Monochromator Grating (SGM) beamline at the Canadian Light Source.\cite{Regier_NIM_2007} For XAS, the fluorescence yield was monitored with an energy-discriminating Princeton Gamma-Tech Sahara silicon drift detector (SDD). Such a detector, with an energy resolution just over 100 eV, allows the separation of fluorescence lines from different edges and different elements. The spectra can then be recorded in partial fluorescence yield (PFY) mode, eliminating possible distortions from additional fluorescence lines. The fluorescence yield technique in general allows us to probe the entire implantation region, as opposed to surface sensitive electron yield techniques.

X-ray Excited Optical Luminescence (XEOL) measurements were performed using the same beamline as for XAS. The XEOL spectra were collected using an Ocean Optics QE65000 spectrometer connected to the measurement chamber using a fiber optic feedthrough. The luminescence photons were collected at approximately a 66$^\circ$ angle from the incident x-ray beam.

The (R)XES experiments were performed using Beamline 8.0.1 at the Advanced Light Source.\cite{Jia_REVSCI_1995} The emission spectra were recorded with a Rowland circle grazing incidence grating spectrometer with spherical gratings. The incident x-rays were 30$^{\circ}$ to the sample surface normal, and the angle between the incident x-rays and the x-rays detected by the spectrometer was 90$^{\circ}$. The linear incident polarization was oriented within the scattering plane.

The approximate depth dependence of the concentrations of implanted ions was simulated using the SRIM program.\cite{SRIM_Manual} Simulations were performed using the same materials and implantation parameters as implemented in experiment. The implantation of a total number of $10^5$ ions for both Mn and Co was simulated in order to obtain an adequately converged profile. Note that SRIM does not consider dynamic changes in the host material during the implantation process, so there is no fluence dependence in the profile shape (other than statistical convergence). Additionally, due to neglected density changes during implantation, SRIM simulations tend to slightly overestimate distribution depths for fluences as large as those used here.\cite{Stepanov_DYNA_2008} However, the SRIM results nonetheless provide a suitable estimate of the implantation depth and ion concentration profiles.

To analyze the Mn and Co $L_{2,3}$ edge x-ray spectra obtained for the samples, single impurity Anderson model (SIAM) calculations were performed.\cite{DeGroot_CLSOS_2008} These calculations include multiplet effects, spin orbit interactions, crystal field splittings, and hybridization with the SiO$_2$ valence band. While XAS spectra are primarily sensitive to oxidation states and local bonding symmetry, RXES spectra exhibit a strong sensitivity to covalent interactions through the presence of ligand-to-metal charge transfer excitations. With the SIAM calculations we can reproduce these excitations (as well as $d$--$d$ excitations), allowing us to determine the degree of interaction between the implanted ions and the host SiO$_2$ matrix. The SIAM calculations are a semi-empirical approach, where we fit the crystal field splitting energy (here $10Dq$ is used), the charge transfer energy ($\Delta$), and the hopping integrals ($V$). The optimal values provide us with information regarding the local coordination of the transition metal ions as well as their hybridization with the host electronic structure. Other parameters, including the intra-atomic Slater integral rescaling factor $\kappa$, the on-site Coulomb repulsion $U$, and the core hole potential $Q$ were fixed at typical values,\cite{DeGroot_CLSOS_2008} with some small refinement employed to optimize agreement with experiment.

For RXES calculations we use the standard Kramers-Heisenberg approach.\cite{Ament_RIXS_RMP_2011} Calculations are performed for the same scattering geometry as experiment, with a 90$^\circ$ scattering angle and linear polarization in the scattering plane. For XAS, we use the same second order scattering formalism, since we used PFY detection in the experiment, but now we integrate the total scattered intensity for each incident energy. Note this give spectra which can be quite different than simple dipole XAS transitions.\cite{DeGroot_FY_SSC_1994} Here the linear incident polarization is in the horizontal plane, with an azimuthal scattering angle of 45$^\circ$ and a detection angle 25$^\circ$ below the plane. 

\section{Results and Discussion}

\subsection{Ion Distributions}

\begin{figure}
\centering
\includegraphics[width = 3.25 in]{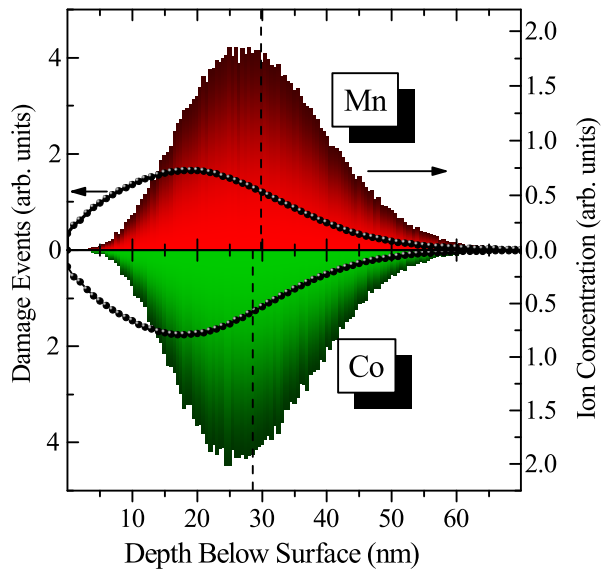}
\caption{(Color online) Implantation depth profiles simulated for the Mn (top) and Co (bottom) ions. The shaded profiles give the distributions of the ions while the dotted line curves denote the damage events. The vertical dashed lines mark the average depths of the implanted ions.}
\label{Fig:Implantation_Depth}
\end{figure}

The simulated depth distributions of the implanted Mn and Co ions are shown using the shaded profiles in Figure \ref{Fig:Implantation_Depth}. Additionally, we plot the simulated depth dependence of damage events (vacancies induced by implanted ions) using the dotted lines. The Mn and Co profiles (both ion concentration and damage events) are very similar, due to the similar masses of the ions. However, Mn is shown to penetrate slightly further into the host on average. This is quantified by the average depths, which are denoted by the vertical dashed lines in the figure. From the simulations, the average depth (or projected range, $R_p$) is 29.8 nm for the Mn ions and 28.5 nm for the Co ions.

As mentioned previously, the profile shapes in Figure \ref{Fig:Implantation_Depth} are independent of implantation fluence, within the SRIM approximations. However, estimates of the average concentrations of the ions can be determined directly from the SRIM results by incorporating the experimental fluence values. For the three experimental doses used, the concentration estimates determined from the SRIM results for volumes contained within two standard deviations of $R_p$ are given in Table \ref{Tab:Concentrations} for both Mn and Co. The differences between the elements are due to both the slightly different implantation profiles, and the slightly different number of ions implanted (recall equal amounts of Mn and Co by weight were used as \mbox{starting materials).} 

\begin{table}
\centering
\caption{\label{Tab:Concentrations}Calculated average concentrations of implanted ions for the fluences used in this work.}
\begin{tabular}{c c c c } \hline \hline
 & \multicolumn{3}{c}{Average Concentration (at. \%)} \\ \cline{2-4}
~~Ion\textbackslash Sample~~ & ~~~~~~L~~~~~~ & ~~~~~~M~~~~~~ & ~~~~~~H~~~~~~ \\   \hline
Mn & 1.15 & 4.74 & 24.89 \\ 
Co & 1.13 & 4.65 & 24.54  \\ \hline \hline
\end{tabular}
\end{table}

\subsection{Coordination of Implanted Ions}

\begin{figure*}
\centering
\includegraphics[width=6.5in]{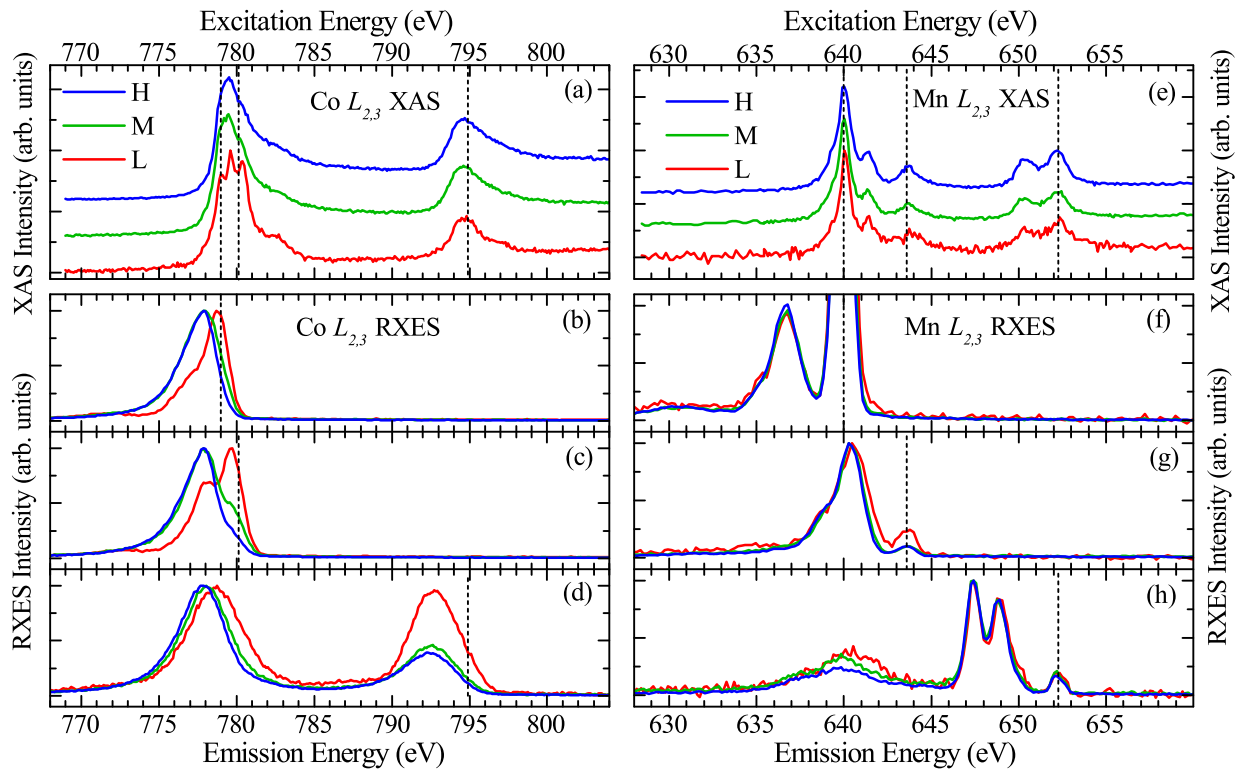}
\caption{(Color online) Co and Mn $L_{2,3}$ spectra for the implanted samples. (a) and (e) show Co and Mn $L_{2,3}$ XAS spectra acquired using partial fluorescence yield, respectively. (b-d) and (f-h) show Co and Mn $L_{2,3}$ RXES spectra on the same energy scale as the corresponding XAS, respectively. The vertical dashed lines denote the RXES excitation energies. Samples are labelled as (H)igh, (M)edium, and (L)ow implantation fluence.}
\label{Fig:CoMn_XAS_XES}
\end{figure*}

To study in detail the specific interactions of the implanted Co and Mn ions with the host lattice, Co and Mn $L_{2,3}$ XAS and RXES measurements were performed and are shown in Figure \ref{Fig:CoMn_XAS_XES}. These measurements provide an element-selective probe of the electronic and crystal structures specific to the implanted ions, and can therefore provide detailed information about the interaction of the ions with the host material.

First, we focus on the Co $L_{2,3}$ XAS and RXES of the implanted samples as shown in the left panels of Figure \ref{Fig:CoMn_XAS_XES}. All spectra are normalized, and the XAS spectra are offset vertically for clarity.  The vertical dashed lines denote the three excitation energies used for the RXES measurements shown in the bottom panels. A clear trend is visible in both the XAS and RXES spectra for different fluences. For the sample irradiated with the highest fluence, \mbox{2$\times$10$^{17}$ cm$^{-2}$} (labelled as H for \emph{high} fluence), the XAS spectrum is broad and featureless, indicative of metallic Co.\cite{Stohr_PRB_2001}  However, with decreasing fluence, the spectra begin to show multiplet features typically present in more ionic materials such as oxides.\cite{Stohr_PRB_2001, deGroot2005}  This same trend is visible in the RXES spectra as well, where the spectra of the high fluence sample appear as fluorescence at a constant emission energy, as expected for metallic systems, whereas those of the lowest fluence sample (L) show sharp $d$--$d$ excitations tracking the excitation energy, again typically present in ionic Co materials such as oxides.\cite{Chiuzbaian_PRB_2008, Magnuson_PRB_2002} Thus, for low fluence implantation, we observe oxygen coordinated Co, while higher fluences show increasing metallic aggregation of Co (likely in the form of Co nanoclusters which have been previously produced in ion implantation studies \cite{Meldrum_AM_2001, Cattaruzza_APL_1998}). Also note that the stark variation in the intensity ratios for the $L_2$ and $L_3$ peaks in panel (d) provide further evidence for a metallic environment for higher fluence and oxygen coordination for lower fluence.\cite{Kurmaev_L23_JELSPEC_2005}

\begin{figure}
\centering
\includegraphics[width=3.25in]{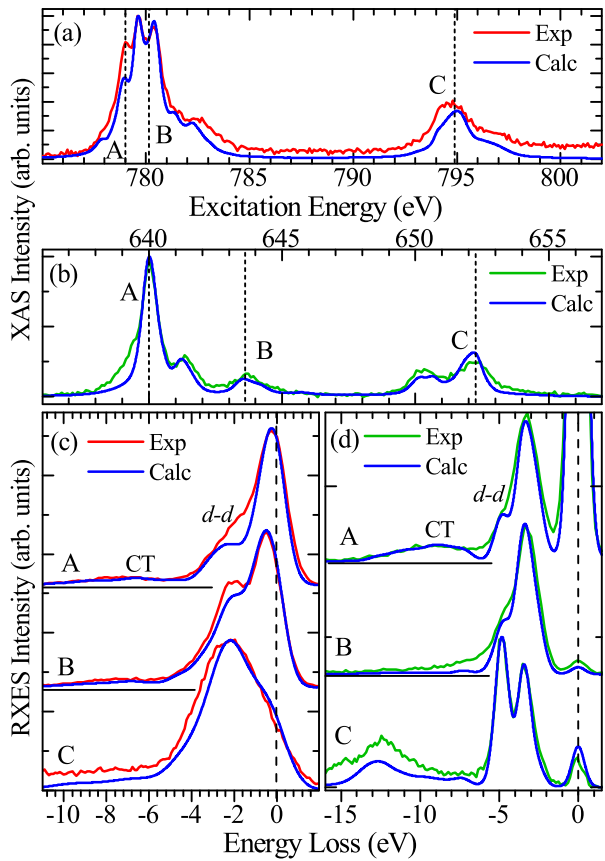}
\caption{(Color online) Experimental Co and Mn spectra, along with SIAM calculations. (a) Co $L_{2,3}$ PFY-XAS. (b) Mn $L_{2,3}$ PFY-XAS. (c) Co $L_{2,3}$ RXES. (d) Mn $L_{2,3}$ RXES.}
\label{Fig:XAS_RXES_Calc}
\end{figure}

Our spectra allow us to investigate in more detail the Co--O interactions for the low fluence sample, by analysis using the SIAM calculations as well as through comparison to spectra of known materials. First, considering the multiplet features of the XAS and the $d$--$d$ excitations of the RXES, it is evident that the Co ions are in a formal 2+ oxidation state with a high spin ground state.\cite{DeGroot_CLSOS_2008} Further, the XAS spectra show a reduced multiplet splitting compared to CoO \cite{DeGroot_CLSOS_2008} (ruling out the formation of CoO-like clusters), and a different shape than that of tetrahedral Co$^{2+}$, found for example in Zn$_{1-x}$Co$_x$O\cite{Krishnamurthy_ZnCoO_JAP_2006} (ruling out significant substitution of Co into tetrahedral Si sites). We find using the SIAM calculations that the Co ions are in a weak, approximately octahedral ligand field, as shown in Figure \ref{Fig:XAS_RXES_Calc}, where we compare calculated and experimental spectra for sample L. The SIAM calculations, using the parameters given in Table \ref{Tab:Calculation_Parameters}, show good agreement with the experiment. Small deviations between calculation and experiment, in particular the onset of the XAS, are consistent with a small portion of metallic Co clustering even in this low fluence sample.

Charge transfer (CT) excitations can be seen in the RXES spectra shown in Figure \ref{Fig:XAS_RXES_Calc}, where we plot the RXES on an energy loss scale to more easily distinguish between $d-d$ and CT features. These CT excitations indicate a significant covalent interaction between the implanted ions and the SiO$_2$, and are due to final states which have a hole in the SiO$_2$ valence band and an extra Co $3d$ electron. Such excitations are captured by the hybridization terms of the SIAM, and the agreement with experiment is very good, again using the parameters of Table \ref{Tab:Calculation_Parameters}. Here we find Co $3d$ -- O $2p$ hopping integral $V_{e_g}$ with a magnitude very similar to what is found in monoxides,\cite{DeGroot_CLSOS_2008, Chiuzbaian_PRB_2008, Magnuson_PRB_2002} indicating a significant covalent interaction between the Co ions and the O atoms of the SiO$_2$. This will be further elucidated when discussing the O $K$ edge results below. Finally, we note that the multiplet features in the $L_3$ region of the XAS at \mytilde 780 eV are very sensitive to the Slater integrals, and optimal agreement was obtained by multiplying the $F^2$ Coulomb integral by an extra factor of 0.9. It is known from optical studies and theoretical considerations that the $F^2$ integral is often screened more than the $F^4$ in transition metal complexes.\cite{Schmidtke_SlaterScreening_2004}

\begin{table}
\centering
\caption{\label{Tab:Calculation_Parameters}Parameters used for the SIAM calculations, as defined in Section \ref{SEC:Exp}. $\kappa$ are given as fractional amounts of the Hartree-Fock values for Slater integrals, while all other values have units of eV. Where two values are given, they are for the non-core-excited and core-excited configurations, respectively. Hopping integral $V_{t_{2g}}$ is fixed at 0.5$V_{e_g}$, and an elliptical band shape 6 eV wide was used for hybridization.}
\begin{tabular}{cccccc} \hline \hline
~~Ion~~ & ~~~~$\kappa$~~~~ & ~~~~$10Dq$~~~~ & ~~$V_{e_g}$~~ & ~~$\Delta$~~ & $U - Q$ \\ \hline
Co$^{2+}$ & 0.80/0.85 & 0.45/0.35 & 1.8 & 5.5 & -0.5 \\
Mn$^{2+}$ & 0.76/0.83 & 0.45/0.35 & 1.8 & 6.6 & -1.0 \\ \hline \hline
\end{tabular}
\end{table}

A similar analysis can be performed for the implanted Mn ions in the samples. In the right panels of Figure \ref{Fig:CoMn_XAS_XES}, we show the experimental Mn $L_{2,3}$ XAS and RXES spectra for all samples, using the same format as the Co data on the left. Unlike the case of Co, all three Mn XAS spectra show strong multiplet features, ruling out the presence of metallic Mn aggregates. As with the Co, the spectra are indicative of Mn in the 2+ oxidation state.\cite{Cramer_Mn_1991} Further, the reduced multiplet splitting in the Mn spectra (i.e. lack of pre-edge multiplet in $L_{3}$ region, and two peaks rather than three in the $L_{2}$ region) indicates that while the Mn is also likely oxygen-coordinated, it is not in the octahedral MnO coordination, but again a rather weaker, possibly distorted coordination similar to the Co.

Consistent with the XAS, we see that the Mn RXES spectra for all samples in panels (f-h) of Figure \ref{Fig:CoMn_XAS_XES} show strong $d-d$ excitations, rather than fluorescence which would dominate for the case of metallic aggregates. There are some small differences in the RXES spectra for the different samples, but these can be attributed to a combination of more statistical noise for the lower fluence samples, and less self absorption (i.e., re-absorption of scattered photons before exiting the sample) in the lower fluence samples.

SIAM calculations are compared to the Mn spectra for sample M in Figure \ref{Fig:XAS_RXES_Calc}. Again we see the assumption of a weak $O_h$ ligand field reproduces the multiplet features of the XAS and RXES spectra very well, with small differences likely due to distortions from $O_h$ in the interstitial sites. Again, as with the case of the low fluence Co results, we see significant CT excitations in the RXES spectra, indicating covalent interactions with the O atoms of the SiO$_2$. Very similar SIAM parameters are found for Mn compared to Co, with the primary difference being the charge transfer energy $\Delta$. This trend is again consistent with differences between Co and Mn monoxides.\cite{Ghiringhelli_MnO_PRB_2006, Chiuzbaian_PRB_2008, Magnuson_PRB_2002} 

\subsection{Effects on Host Electronic Structure}

\begin{figure}
\centering
\includegraphics{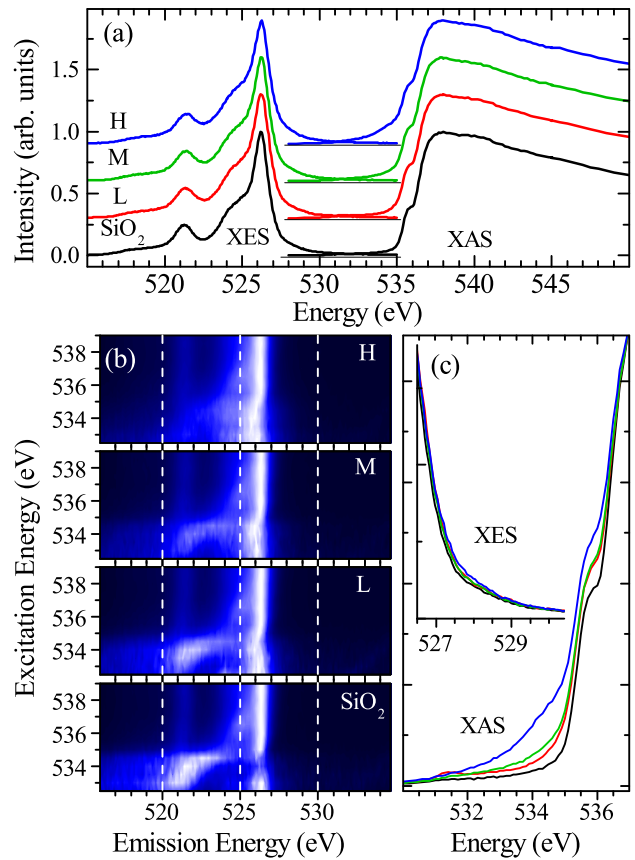}
\caption{(Color online) Oxygen $K$ edge spectra. (a) XES is shown on the left, measured with a 560 eV incident energy, and XAS is on the right. (b) Resonant XES (RXES). (c) Enlarged view of the near-gap XAS and XES (inset).}
\label{Fig:Oxygen_XAS_XES}
\end{figure}

The XAS and XES at the oxygen $K$ edge are shown in the upper panel of Figure \ref{Fig:Oxygen_XAS_XES}. The spectra exhibit the general shape expected and observed previously for SiO$_2$.\cite{Green_SiO2PbSn_2012}  The most prominent changes evident in the spectra upon implantation are in the pre-edge region of the XAS. In the range of \mytilde 532-535 eV, as expanded in the lower right panel of the figure, there is increasing spectral weight for increasing implantation dose.  It is known that this region of the oxygen $K$ edge XAS for transition metal (TM) oxides reflects the O $2p$ to TM $3d$ hybridization.\cite{deGroot_PRB_1989} Thus, here we see further evidence of bonding between the Co and Mn with the O atoms in the SiO$_{2}$ matrix, as was detected above via the SIAM analysis of the $L$ edge. Now we further see that due to the location of $2p-3d$ states at the low energy threshold of the conduction band, their introduction leads to a reduction in the overall band gap of the material.

The XES spectra in the upper panel of Figure \ref{Fig:Oxygen_XAS_XES}, unlike the XAS, show very minimal changes upon implantation, even if overlaid as in the lower right panel of the figure.  This is likely due to the fact that the occupied Co and Mn $3d$ states are deeper within the valence band, and therefore do not act to raise the VB maximum. We can better identify these deeper states using RXES spectra. The lower left panels of Figure \ref{Fig:Oxygen_XAS_XES} are spectral maps displaying resonantly excited XES (RXES) for the samples.  The excitation energies span from 532 to 539 eV, covering the part of the XAS which shows increased presence of transition metal $3d$ states. Note that the emission energy scale for these maps is aligned with that of the top panel, so that the two main peaks in the XES spectrum line up for all panels. Here, the influence of the implanted ions on the valence band structure is more evident. The spectra show increasing weight at \mytilde 525 eV emission energy for increased implantation dose.  This is magnified for an excitation energy of 534 eV, where SiO$_2$ shows almost no weight at 525 eV emission, while the implanted samples have significant weight. In addition to the gradual introduction of Mn and Co occupied $3d$ states, the gradual smearing of the spectra for increasing implantation dose is also possibly due to damage of the host structure as a consequence of the implantation process. With higher fluences, we have higher damage which can lead to a smearing of the electronic structure.

\begin{figure}
\centering
\includegraphics[width=3.25in]{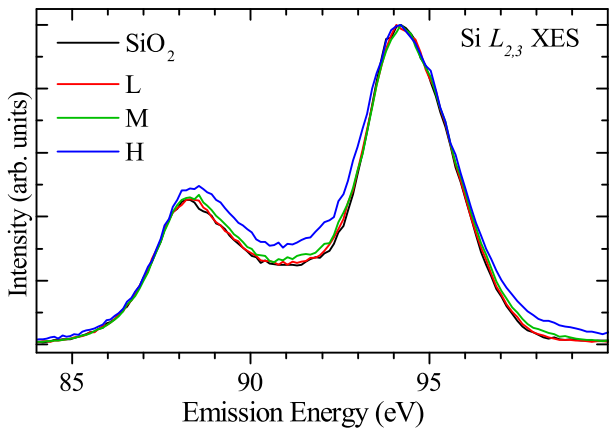}
\caption{(Color online) Silicon $L_{2,3}$ XES spectra (measured with 120 eV incident energy) for the implanted samples and reference SiO$_2$ sample.}
\label{Fig:Silicon_XAS_XES}
\end{figure}

In Figure \ref{Fig:Silicon_XAS_XES}, we show XES spectra taken at the Si $L_{2,3}$ edge for all samples. The spectra exhibit the general shape expected for SiO$_2$.\cite{Sham_PRB_2004} However, we again see a trend among the normalized spectra. As the implantation fluence increases, there is increased spectral weight between the main peaks (from \mytilde 88 -- 94 eV), as well as at higher energies (97 -- 99 eV). Such spectral changes have been observed previously in ion-implanted SiO$_2$ and can be ascribed to the formation of Si-Si bonds, which can arise due to the creation of O vacancies during the implantation process.\cite{Green_SiO2Mn_JPCM_2012} With higher fluence one expects more damage, which leads to an increasing concentration of pure Si regions as indicated by the trend in the Si $L$ edge spectra.

\subsection{Luminescence}

\begin{figure}
\centering
\includegraphics[width=3.25in]{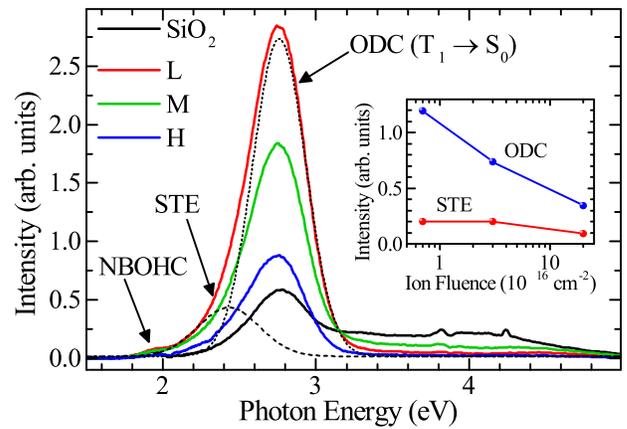}
\caption{(Color online) X-ray Excited Optical Luminescence spectra for the implanted and reference samples. Dashed lines show Gaussian profiles fitted to the spectrum of sample L. The inset shows the change in intensity of the self trapped exciton (STE) and oxygen deficient center (ODC) components for varying implantation fluence.}
\label{Fig:XEOL_Plot}
\end{figure}

In Figure \ref{Fig:XEOL_Plot}, we show x-ray excited optical luminescence (XEOL) spectra for the implanted and reference samples.  An excitation energy of 520 eV was used, with the incident X-rays at an angle of 5$^{\circ}$ to the surface normal and the XEOL detector was at an angle of 66$^{\circ}$ to the incident radiation. First, we see the non-implanted sample shows weak, broad luminescence. X-ray excited luminescence over this band from 2.5 -- 5 eV in quartz has previously been observed, and attributed to defects such as impurities and self-trapped excitons.\cite{Sham_QuartzXEOL_PhysChemMin_2009} Such processes appear to be the cause of the majority of luminescence for the SiO$_2$ sample here. The reference sample also shows a weak feature at \mytilde 1.9 eV, which has been attributed to dangling oxygen bonds termed non-bridging oxygen hole centers (NBOHC).\cite{Skuja_ODC_NBOHC_JNCS_1998,Trukhin_JNCS_STE_NBOHC_1998,Kajihara_NBOHC_APL_2001}

Upon implantation, the luminescence changes drastically, and a very intense line centered at 2.75 eV is observed for the lowest fluence sample. The overall shape can be deconvolved into two Gaussian profiles centered at \mytilde 2.75 eV and \mytilde 2.4 eV. The more intense peak at 2.75 eV is identical to that of the well-known triplet luminescence for oxygen-deficient centers (ODC),\cite{Skuja_ODC_NBOHC_JNCS_1998,Trukhin_JNCS_STE_NBOHC_1998,ZatsepinAF_ODC_PSS_2010} while the weaker peak is similar to silica peaks attributed to self-trapped excitons.\cite{Trukhin_JNCS_STE_NBOHC_1998,ZatsepinAF_STE_JLum_2013} For increasing implantation fluence we see that the same general shape is present, but the overall intensities of the peaks decrease, as shown qualitatively in the main panel of Figure \ref{Fig:XEOL_Plot} and quantitatively by the fitted Gaussian intensities in the inset.

There are several different model approaches for characterizing the local atomic structure of ODCs.\cite{Skuja_ODC_NBOHC_JNCS_1998} One of them assumes the ODC is a neutral oxygen vacancy, i.e. a non-regular Si-Si bond.\cite{ZatsepinAF_ODC_PSS_2010,ZatsepinAF_STE_JLum_2013} The ODC band in our data is consistent with the \mbox{T$_1 \rightarrow$ S$_0$} radiative transition. Such a situation usually can be realized under optical excitation due to intercombined conversion of the ODC excited states \cite{ZatsepinAF_ODC_PSS_2010,Skuja_ODC_NBOHC_JNCS_1998} and at room temperatures results in dominant triplet luminescence (the singlet ODC luminescence at 300 K is usually quenched). The decrease in luminescence intensity with increasing implantation fluence seems to arise from a quenching due to the Co and Mn ions, which can lead to effective non-radiative recombination of charge carriers on defect centers. Thus, it seems the most active luminescence might be achieved by a slightly lower fluence, where the production of ODCs is balanced with the quenching due to Co and Mn. These results suggest that this dual simultaneous implantation has a wide range of possibilities for transforming and regulating the optical properties and electron energy spectrum (electronic structure) of  advanced materials that may be suitable for optoelectronics and nanophotonics.

\section{Conclusion}

We have studied the implantation of Mn and Co ions into amorphous SiO$_2$. We find that at lower implantation fluences the Co ions occupy interstitial locations in the host, whereas the Co aggregates into metallic clusters for larger fluence. Conversely, the Mn ions remain in interstitial locations for all fluences studied. We find that the implantation of these ions introduces new conduction band states which act to lower the electronic band gap of the material, while occupied states are deeper in the valence band and have less effect. Finally, the implantation process using a low fluence greatly enhances the x-ray excited luminescence for a band of wavelengths centered at \mbox{2.75 eV.}

\begin{acknowledgments}
This work was supported by the Natural Sciences and Engineering Research Council of Canada (NSERC), the Canada Research Chair Program, the Ural Division of the Russian Academy of Sciences (Project No. 12-I-2-2040), and the Russian Science Foundation for Basic Research (Project Nos. 13-08-00059, 13-08-00568, and 13-02-91333). We gratefully acknowledge the assistance from the staffs of the Canadian Light Source at the University of Saskatchewan and the Advanced Light Source at Lawrence Berkeley National Laboratory. The Advanced Light Source is supported by the Director, Office of Science, Office of Basic Energy Sciences, of the U. S. Department of Energy under Contract No. DEAC02-05CH11231. The Canadian Light Source is supported by NSERC, the National Research Council (NSC) Canada, the Canadian Institute of Health Research (CIHR), the Province of Saskatchewan, Western Economic Diversification Canada, and the University of Saskatchewan.
\end{acknowledgments}

\end{document}